\documentclass[journal,twocolumn,letterpaper]{IEEEJERM}

\usepackage{url}
\usepackage{multirow}
\usepackage{times,amsmath,epsfig}
\usepackage{fancyhdr}
\usepackage{amsmath}
\usepackage{amsfonts}
\usepackage{amssymb}
\usepackage[utf8]{inputenc}
\begin{document}

%

\title{CV-Attention UNet: Attention-based UNet for 3D Cerebrovascular Segmentation of Enhanced TOF-MRA Images}

\author{Syed Farhan Abbas$^1$, Nguyen Thanh Duc$^1$, Yoonguu Song$^1$, Kyungwon Kim$^1$, Ekta Srivastava$^2$, Boreom Lee$^*$~\IEEEmembership{Member, IEEE}
}


\twocolumn[
\begin{@twocolumnfalse}
\maketitle 


\begin{abstract}
 
Background: \textnormal{Due to the lack of automated methods, to diagnose cerebrovascular disease, time-of-flight magnetic resonance angiography (TOF-MRA) is assessed visually, making it time-consuming. The commonly used encoder-decoder architectures for cerebrovascular segmentation utilize redundant features, eventually leading to the extraction of low-level features multiple times.  Additionally, convolutional neural networks (CNNs) suffer from performance degradation when the batch size is small, and deeper networks experience the vanishing gradient problem.} Methods: \textnormal{In this paper, we attempt to solve these limitations and propose the 3D cerebrovascular attention UNet method, named CV-AttentionUNet, for precise extraction of brain vessel images. We proposed a sequence of preprocessing techniques followed by deeply supervised UNet to improve the accuracy of segmentation of the brain vessels leading to a stroke. To combine the low and high semantics, we applied the attention mechanism. This mechanism focuses on relevant associations and neglects irrelevant anatomical information. Furthermore, the inclusion of deep supervision incorporates different levels of features that prove to be beneficial for network convergence.} Results: \textnormal{We demonstrate the efficiency of the proposed method by cross-validating with an unlabeled dataset, which was further labeled by us. We believe that the novelty of this algorithm lies in its ability to perform well on both labeled and unlabeled data with image processing-based enhancement. The results indicate that our method performed better than the existing state-of-the-art methods on the TubeTK dataset.} Conclusion: \textnormal{The proposed method will help in accurate segmentation of cerebrovascular structure leading to stroke.} 

\end{abstract}

\begin{IEEEkeywords}
Attention mechanism, cerebrovascular segmentation, deep supervision, time-of-flight magnetic resonance angiography (TOF-MRA), 3D-UNet.
Clinical and Translation Impact: \textnormal{: The proposed method can be beneficial for assessment of cerebrovascular structures and diseases that are relating to diagnosis and prognosis of stroke.} 
\end{IEEEkeywords}

\end{@twocolumnfalse}]

{
  \renewcommand{\thefootnote}{}
  \footnotetext[1]{ This work was supported by GIST Research Institute (GRI) ARI grant funded by the GIST in 2021. This work was also supported by the Technology Innovation Program (Industrial Strategic Technology Development Program-Development of Core Industrial Technology) (20003822, Development of Navigation System Technologies of MicroNano Robots with Drug for Brain Disease Therapy) and funded by the Ministry of Trade, Industry \& Energy (MOTIE), Korea.}
  \footnotetext[2]{Corresponding Author: $^*$Boreom Lee is with Department of Biomedical Science and Engineering, Gwangju Institute of Science and Technology (GIST), Gwangju, 61005, South Korea (email:leebr@gist.ac.kr).}
  \footnotetext[3]{$^1$Syed Farhan Abbas, $^1$Yoonguu Song, $^1$Nguyen Thanh Duc, $^1$Kyungwon Kim was student of Department of Biomedical Science and Engineering, Gwangju Institute of Science and Technology, Gwangju 61005, Republic of Korea (e-mail: farhanabbas@gm.gist.ac.kr).}
  \footnotetext[4]{$^2$Ekta Srivastava is with Department of Electrical Engineering, Indian Institute of Technology (IIT), Dehli, India.}
}
 
%
\IEEEpeerreviewmaketitle

\section{Introduction}
\label{sec:introduction}

\IEEEPARstart{S}{troke} is a medical condition related to the cerebrovascular system. Any diseases that are present in the body before stroke, therefore, have a great impact on people brain; thus, knowing the vascular anatomy is critical to neurosurgeons and affects the overall healthcare system, as stroke is ranked the second leading cause of death in the world \cite{donkor2018stroke}. Ischemic stroke, along with several cerebrovascular diseases, such as aneurysms, arteriovenous malformations, carotid stenosis, and vessel occlusion, are all related to interruption in blood supply and vessels \cite{wang2015threshold}. For early diagnosis of cerebrovascular diseases, having accurate details of the cerebrovascular structure is of great importance for treatment and diagnosis of chronic cerebrovascular disease. In clinical practice, the segmentation of brain vessels can predict stroke events and can provide pivotal information before surgery, aiding in the diagnosis of abnormalities.

Analysis of blood vessels is challenging due to their size, overlap with other blood vessels, contrast with anatomical structures, and tortuosity. Widely used noninvasive imaging modalities for cerebrovascular disease research include computed tomography (CT), magnetic resonance imaging (MRI), positron emission tomography (PET), and X-ray. Using these modalities, radiologists and surgeons can visualize brain lesions, tissues, or blood vessels. In MRI, angiographic images are called magnetic resonance angiography (MRA). Time-of-flight MRA (TOF-MRA) is a non-contrast enhanced MRA that is a commonly used modality for the visualization of blood vessels. TOF-MRA is suitable for depicting intracranial arteries, as it is less susceptible to intravoxel dephasing and has a high spatial resolution without the need to inject a contrast agent \cite{kim20123d}. However, in slow blood flow and small vessel areas, the imaging quality is quite poor because of the bias field and noise. Furthermore, the clinician manually looks for abnormalities in the brain scans because no automated technique for analyzing these brain vessels exists. A fully automated system requires brain vessel segmentation without post processing on the scanner console. A hurdle in applying the clinical application of cerebrovascular segmentation widely in practice is the time consumed and non-standardized post image processing. Therefore, end-to-end analysis of TOF-MRA can quantify the cerebrovascular status and is beneficial for the diagnosis of vessel abnormalities.

For fully automated systems, researchers use deep learning models for the diagnosis of different abnormalities in various organs. Many convolutional neural networks (CNNs) are fully automated and contain encoder-decoder modules with two-dimensional (2D) data, which require a large, labeled dataset to train the model. Furthermore, in most medical image segmentation problems, the affected area is cluttered or centered in one area, for example, a tumor, lesion, or organ segmentation. However, vessels are tubular in structure and spread across organs with varying diameters and lengths. This topological structure in the brain makes the segmentation task complex. Hence, the labeling of 3D TOF-MRA data is a time consuming and an intricate task.

Some approaches have been applied to segment brain vessels automatically using traditional approaches along with deep learning methods. However, these approaches usually require a large amount of labeled 2D data. It is worth mentioning that 3D cerebrovascular segmentation has not been studied extensively, and there is room for improvement, as the 3D approach can obtain more contextual information with less data than 2D-based approaches. Owing to the complex geometry and imbalanced illumination of small and large vessels, the existing methods either over-segment or miss-segment the vessels. To fill this gap, it is necessary to design a 3D model that provides fully automated segmentation.

In this paper, we propose a 3D end-to-end cerebrovascular segmentation method that can outperform state-of-the-art approaches on 3D TOF-MRA data. First, we included the vessel enhancement method to illuminate small and large vessels and reduce complex background anatomical structures. Second, we apply the novel attention mechanism-based 3D-UNet to cerebrovascular segmentation. Finally, with the enhanced attention-based 3D-UNet, we achieved better performance in terms of several segmentation evaluation metrics for 3D TOF-MRA data.

This paper contains five sections and is structured as follows:  Section \ref{sec:related works} presents works related to the model driven and data-driven approaches, along with vessel enhancement techniques for vessel segmentation. Section \ref{Method} explains our proposed methodology. Section \ref{results} presents the experimental results and provides a discussion of the TubeTk dataset. Section \ref{conclusion} concludes the paper.

\section{Related Works}
\label{sec:related works}
Two popular approaches for cerebrovascular segmentation aimed at helping radiologists and neurosurgeons provide diagnoses are the model-driven and data-driven approaches, which are described in detail below.

\subsection{Model Driven Approaches }
Model-driven approaches for cerebrovascular segmentation have been studied widely based on region, active contour, statistics, and multiscaling. Region-based approaches find certain similarities in the image and segment it. Region-based approaches are divided into two categories centered on threshold and growth regions. Furthermore, threshold-based approaches \cite{babin2011segmentation} divide the region into soft tissue, vessels, background, and bone structure, while region-growing studies \cite{wang2015threshold} build arterial and venous trees by considering the neighboring voxels derived from the initial seed point or initial selected grayscale value. In active contours or deformable model-based techniques \cite{lorigo2001curves}, the boundary of a blood vessel image is estimated using the energy minimization connectivity preserving method. The geodesic active contour based on external and internal forces forms a closed surface using the minimization of the object \cite{zhao2015extraction}. Similarly, geometrically active contours solve differential equations, called level set segmentation, and this technique can preserve the topology and shape of blood vessels \cite{wang1999improving}. Statistical approaches use optimization algorithms to calculate the probability of a blood vessel. Some approaches include the expectation-maximization (EM) method to classify voxels \cite{zhang2008medical}. For instance, normal and Rayleigh distribution-based methods are used for vessel and nonvessel modeling \cite{hassouna2006cerebrovascular}, while statistical intensity-based approaches \cite{wen2015novel} and other methods normally use the EM algorithm for the estimation of vessels \cite{gao2011fast}. As the vessel width varies along the length, obtaining information at different scales can be beneficial and can be obtained by using the multiscale approach \cite{lu2016vessel}, which can be used with multi-modality angiography images [13]. Therefore, the model-based approach cannot be automated, as it is overwhelmed by handcrafted features, requires many hyperparameters for tuning and is less robust to distinguishing microvessels.

\subsection{Data Driven Approaches}
Recently, deep learning algorithms, specifically CNNs, have been driving progress in many fields, specifically in visual recognition tasks, because of their ability to extract nonlinear features. The CNN-based model is the predominant model for overcoming the medical image segmentation problems in brain \cite{dolz2018hyperdense}, cardiac \cite{bernard2018deep}, abdominal CT \cite{roth2018spatial}, as well as lung nodule detection. The most famous CNN model for medical image segmentation is inspired by the fully convolutional neural network (FCN) \cite{long2015fully}, called UNet. It contains a contracting path that extracts high-level features and expands the path for reconstructing the pixelwise segmentation mask. However, in this approach, many low-level features are extracted multiple times, which may be inappropriate for complex medical imaging tasks. Therefore, to encounter and solve complex problems, researchers have proposed many variants of UNet \cite{Xnetnewadd}, \cite{Dnetnewadd}, \cite{zhou2018unet++}.

\subsubsection{Attention based mechanisms }
As described earlier, the encoder-decoder strategy leads to excessive and redundant use of resources. Therefore, an alternative solution, i.e., attention gates (AGs), was proposed, which can highlight salient features and focus on target structures. AGs have been commonly used in natural image analysis \cite{wang2017residual} and classification. Some studies have proposed methodologies such as attention modules in multiscale resolution used for prostate segmentation on ultrasound images \cite{wang2018deep}, hierarchical aggregation attention for left atrial segmentation \cite{chen20173d} and pancreas segmentation based on spatial attention \cite{oktay2018attention}.

\subsubsection{Deep learning approaches for cerebrovascular segmentation}
In cerebrovascular segmentation, researchers have applied CNN-based methodologies. In a recent work, \cite{phellan2017vascular} used a simple CNN for blood vessel segmentation, but the algorithm was not able to segment small vessels. \cite{tetteh2020deepvesselnet} trained the model on synthetic vascular data, which gave promising results on synthetic and clinical TOF-MRA data, but in some areas, the results were over and under-segmented. \cite{zhang2020cerebrovascular_1} applied statistical methods to generate labels and dilate dense convolution and attained quite good Dice similarity coefficient (DSC) scores; however, their label generation method was quite complex. Moreover, their method could not differentiate between arteries and veins, which is important in clinical settings. Y-net \cite{chen20173d} is a similar approach to 3D-UNet \cite{cciccek20163d} but has vessel discontinuity at bifurcation. The 2D patch-based approach is used in \cite{livne2019u}. They applied 2D-UNet, but their approach was not end-to-end and did not consider the contextual information of 3D vessel structures. Most recently, a method called the Uception model was proposed based on combining 3D-UNet and inception modules \cite{phellan2017vascular}, but it was limiting to only macrovessel segmentation.

The CS$^2$-Net studied in \cite{mou2020cs2} was able to design an attention based cerebrovascular segmentation network but could only segment the Circle of Willis. Another network based on residual blocks and reverse edge attention called RE-NET \cite{zhang2020cerebrovascular} performed cerebrovascular segmentation, but their approach still had small vessel discontinuity.

\begin{figure}[]
  \centering
  \includegraphics[width=9cm, height= 4cm]{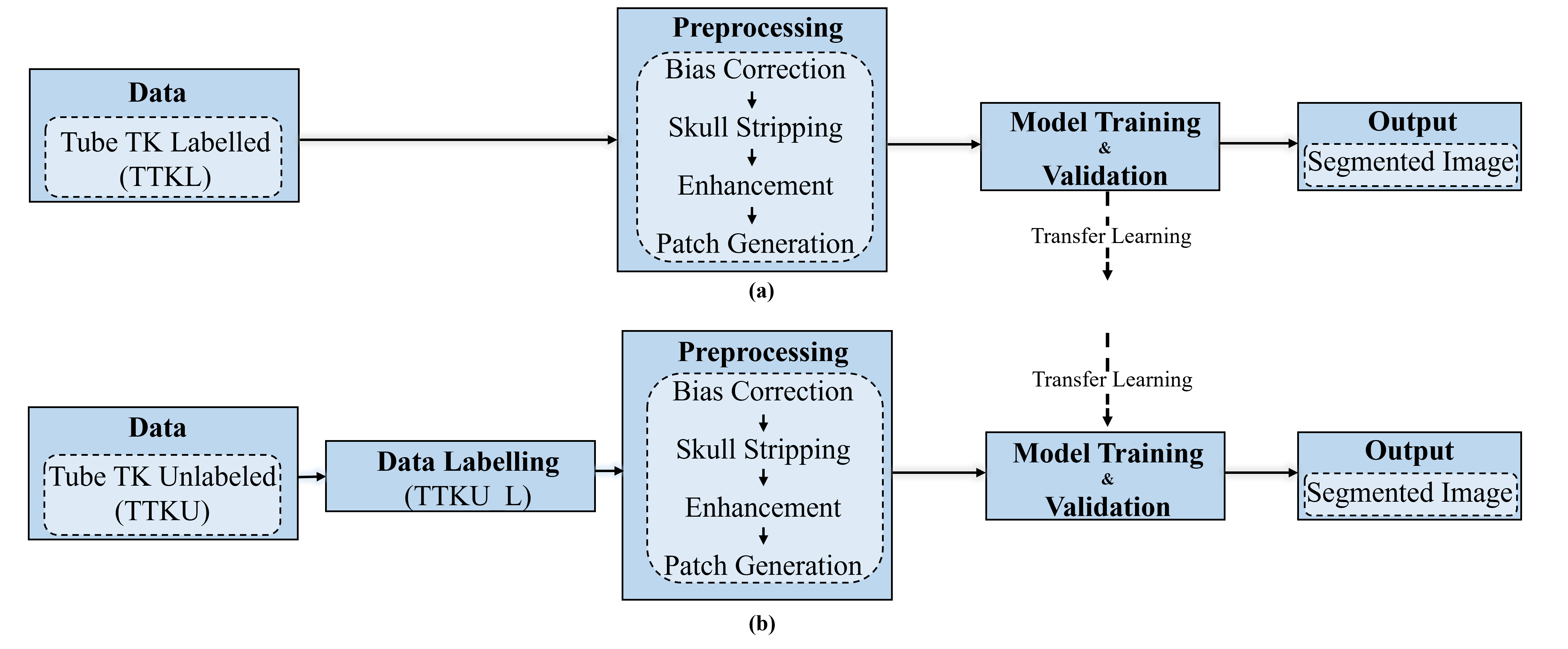}
  \caption {The overall diagram for 3D cerebrovascular segmentation using the deep learning model.}
   \label{Figure:1}
\end{figure}
\begin{figure*}[]
  \centering

  \centerline{\includegraphics[width=\textwidth]{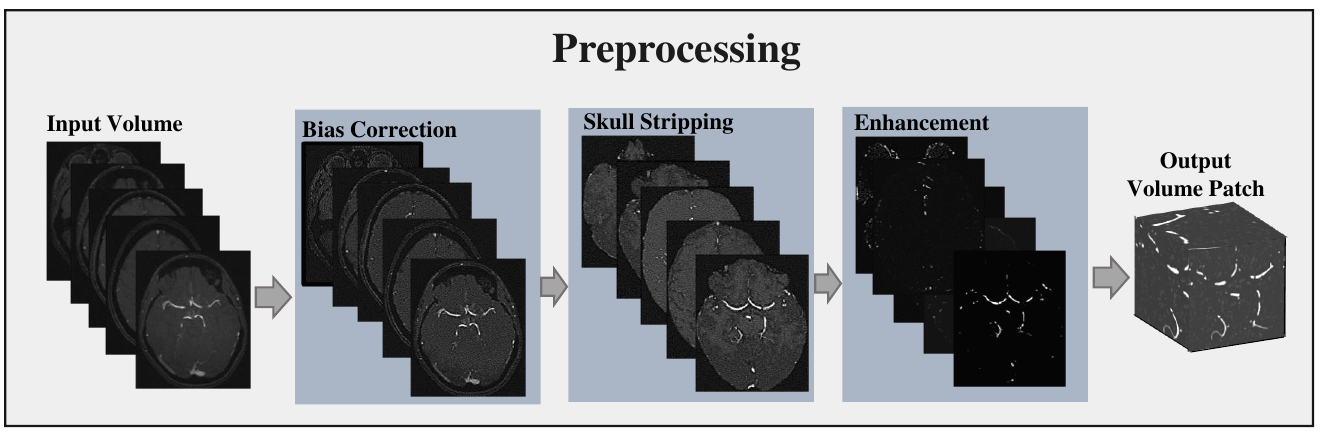}}
  \caption {Preprocessing steps: From left to right, the steps are bias correction, skull stripping, Hessian based vessel enhancement and 3D patch generation.}
  \label{Figure:2}
 
\end{figure*}
To our understanding, there are many challenges in cerebrovascular segmentation of TOF-MRA images. Some difficulties are due to the MRA modality, while some are model generalization deficiencies of imbalanced data, such as vessels. TOF-MRA images have low overall contrast, weak contrast between the background anatomical structure and vessels, smaller vessel sizes and easily deformable tubular vessel shapes. To encounter these issues, we explored the vessel enhancement method using Hessian-based vessel enhancement filtering as a preprocessing step. 3D patches prove advantageous in increasing training samples, using full spatial information, and reducing the network scale, making them computationally efficient while saving memory.

In light of the above mentioned limitations and alternatives, we proposed a 3D cerebrovascular segmentation end-to-end attention-based method for use on enhanced TOF-MRA images, called cerebrovascular attention UNet (CV-AttentionUNET). A UNet-based framework is used to learn the effective 3D features, and an attention module is applied to look for vessels. To our knowledge, this is the first attempt to include attention and enhancement, which has proven to be effective in \cite{phellan2017comparison} for cerebrovascular structures and will help surgeons and radiologists better diagnose cerebrovascular diseases.
\section{Methodology}
\label{Method}
\subsection{Overview of our method }
In this section, we present our proposed methodology for cerebrovascular segmentation, as shown in  Fig. \ref{Figure:1} (a) \& (b). We considered two different datasets comprising labeled and unlabeled images, TTKL and TTKU. Both datasets were subjected to preprocessing steps, including vessel enhancement, and 3D input patches were fed to the network. The network is based on 3D-UNet, which uses spatial attention and deep supervision, named cerebrovascular attention UNet (CV-AttentionUNet), to obtain precise cerebrovascular segmentation. The designed methodology can take 3D images as input and produce 3D vessel-segmented images as an output. The novelty of this algorithm lies in its ability to perform well on both TTKL and TTKU with image processing-based enhancement.

\subsection{Data labelling}
As the data size is small, we labeled the unlabeled TOF-MRA images in the TTKU dataset, resulting in TTKU\_L. We labeled the Circle of Willis and three major brain arteries as follows: the vertebral arteries (VA), internal carotid arteries (ICA), and the basilar artery (BA). These arteries are responsible for carrying most of the blood. We performed prelabeling using a region-growing algorithm implemented in open source MeVisLab \url{https://www.mevislab.de} and corrected by an experienced clinician (coauthor: Kyungwon Kim). To avoid inter-rater variability in the prelabels, it was further checked by a clinician (corresponding author: Boreom Lee), and then the final ground truth was obtained. The average time for labeling TOF-MRA images per person is 30–40 minutes.

\subsection{Preprocessing}
There is nonuniformity in the illumination of brain images, which is called the bias field. To reduce the data inconsistency and nonuniformity, a nonparametric bias field correction is applied. Subsequently, skull stripping is applied to obtain full brain masks, followed by the enhancement method and dividing the whole MRA into 3D patches, as depicted in Fig. \ref{Figure:2}; all preprocessing steps are performed using the ITK library \cite{mccormick2014itk}.

\subsubsection{Enhancement}
Owing to the complexity of vessels and the nonuniform contrast, enhancement methods have been used in retinal vessels \cite{miri2009comparison}, hepatic vessels \cite{luu2012evaluation}, airways \cite{meng2017airway}, and other tubular structure segmentations. To overcome the low contrast and non-uniformity of contrast between larger and smaller vessels in MRA, we applied the Hessian matrix-based Frangi vesselness filter (HFV) over TTKL and TTKU\_L, as shown in Fig. \ref{Figure:1} (a) and Fig. \ref{Figure:1} (b). We used HFV because of the better DSC score and robustness to noise in 3D images compared to other filters \cite{phellan2017comparison}. The HFV filter extracts the vessels from a 3D MRA image by classifying the eigenvalues via the vesselness function of the Hessian matrix. The reconstructed image is observed at a particular scale by convolution of the normal kernel with standard deviation, and a normalized Hessian matrix is generated at a particular scale. Then, a 3D vesselness filter is applied over processed volumes, which considers the dissimilarity of blob-like, plate-like, and line like patterns. To reduce the background structure, the norm of the Hessian matrix is defined and applied as a filter at multiple scales so that the maximum response is obtained as the final output. A mathematical description is presented in \cite{frangi1998multiscale}.

\subsection{CV-AttentionUNet Architecture}
\begin{figure*}[]
  \centering

  \centerline{\includegraphics[width=\textwidth]{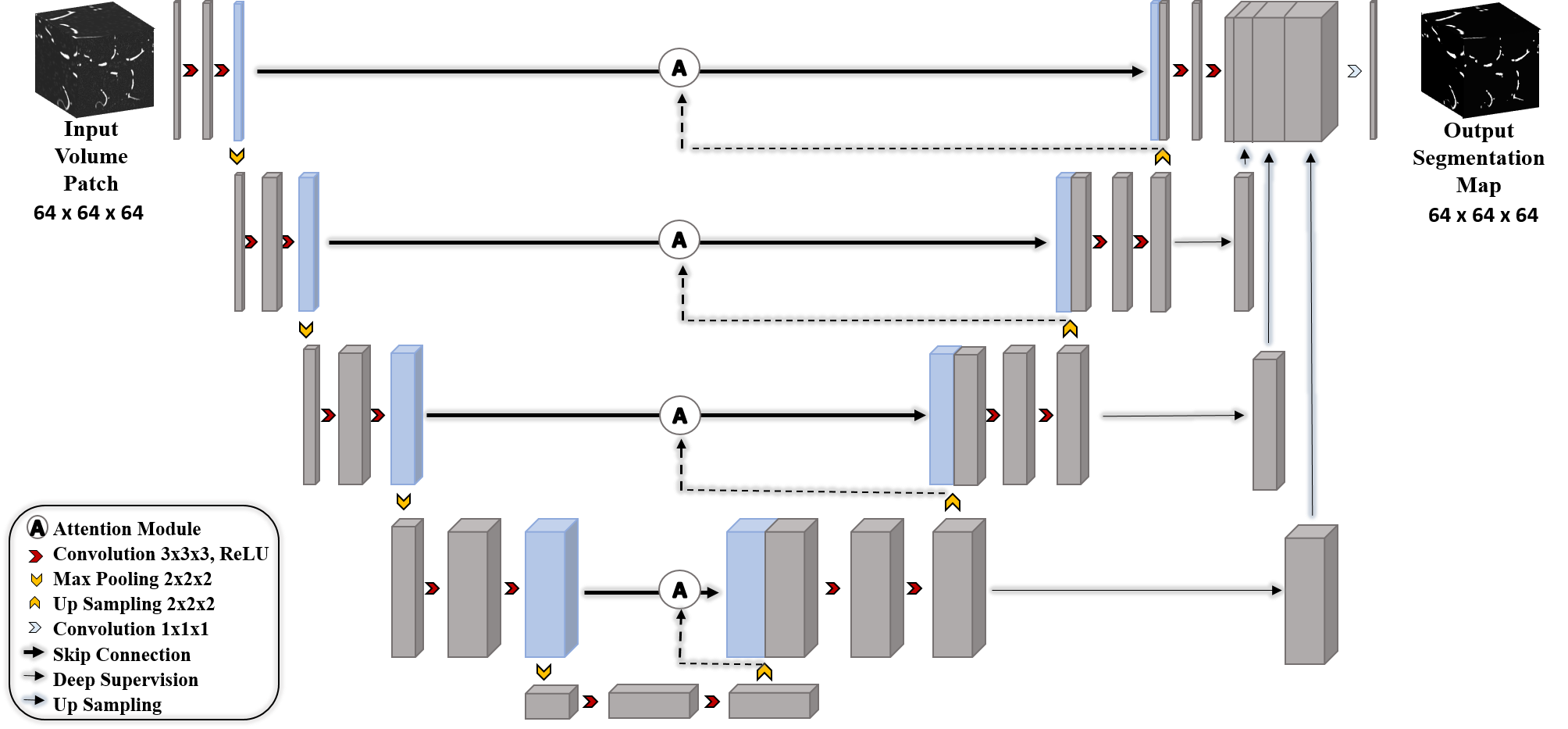}}
  \caption {Proposed cerebrovascular attention UNet architecture (CV-AttentionUNet) and deep supervision.}  \label{Figure:3}\medskip
\end{figure*}

Based on the attention mechanism for medical image segmentation described in \cite{oktay2018attention}, we proposed CV-AttentionUNet, in which the attention mechanism detects vessels while considering the smaller batch size problem. Deep supervision was further utilized to use the features in the intermediate layers. An overview of the architecture is shown in Fig. \ref{Figure:3}.

\subsubsection{3D-UNet as backbone architecture and its limitations }
UNet is well known for performing semantic tasks, and the core idea is to gradually fuse the low semantics with fine spatial maps from the encoder to the high semantic but coarse feature maps from the decoder. Specifically, the encoding path contains a convolutional network with five levels, and each level constitutes two padded convolutions ($7\times7\times7$, $3\times3\times3$), i.e., a batch normalization (BN) layer and a rectified linear unit (ReLU). At each level, a pooling operation ($2\times2\times2$) is applied to downsample the data and to avoid overfitting. The number of feature channels is doubled for accurate and deep feature extraction. In the decoder path, the pooling operations are replaced by upsampled operations ($2\times2\times2$), followed by the same padded convolutions BN and ReLU; additionally, the number of feature channels is halved at every decoding path. In each step, there is a skip connection from each encoder level to the same decoder level to restore the resolution of fine-grained details of the image in terms of gradient flow and to avoid the vanishing gradient problem \cite{drozdzal2016deep}.

However, UNet is not an optimal solution for semantic segmentation, and there still exist limitations; thus, more investigation is needed. First, the skip connections from encoder to decoder, which are at multiple scales, seem to use low-level features. If the low-level encoder features have rich semantic details and fusion methods can concentrate on salient spatial regions of the encoded features, it can obtain more enriched and fine-grained semantic details. Second, there are semantic gaps between the encoder and decoder features, and concatenating the two incompatible features results in poor voxelwise probability maps. Therefore, information sharing between the encoder and decoder could be more effective if it is able to focus on important features and ignore irrelevant features.

\subsubsection{Attention Gate for vessels }
Inspired by the abovementioned shortcomings of UNet, we proposed a spatial attention gate that searches for relevant activations and preserves them for specific tasks. The attention gate generates the attention coefficient $\alpha$, which can identify salient regions and keep the feature responses that are relevant. Feature map (g) and feature map (f) in the decoder and encoder, respectively, are the two inputs of the attention gate. The feature map acts as a gating signal and contains more contextual information regarding low-level features, as suggested in \cite{wang2017residual}. The linear transformation of both vectors is performed after a convolution of kernel size of $1\times1\times1$, a stride window of $1\times1\times1$1, and group normalization (GN), which is also called vector concatenation-based attention.

\begin{figure}[]
  \centering
  \centerline{\includegraphics[width=8.5cm]{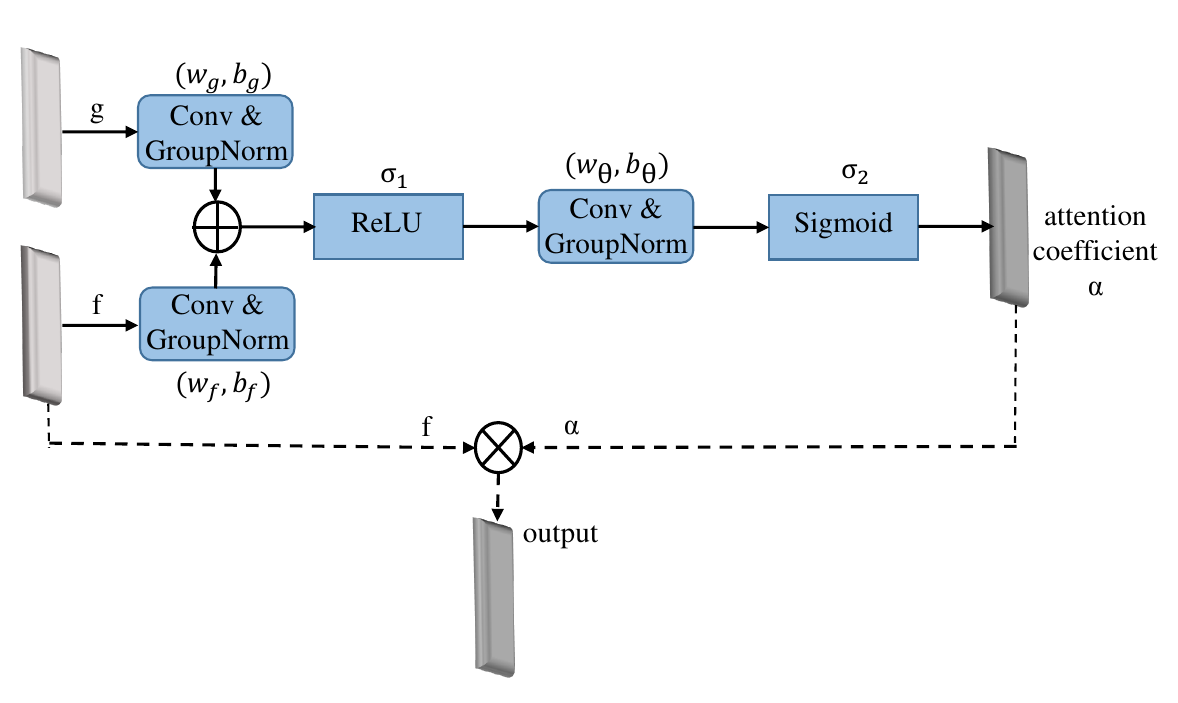}}
  \caption {Diagram for attention module.}\label{Figure:4}\medskip
\end{figure}

Here, we have to emphasize that GN is independent of batch size. We have to use larger patch sizes to obtain rich contextual information, but that batch size still needs to be small due to computational resources limitations. However, as suggested in \cite{wu2018group}, the validation error greatly differs with smaller batch sizes with the use of BN, and training error curves also fluctuate with the use of large to small batch sizes. On this basis, we have used GN in our whole architecture, which boosts the overall performance of the architecture. The vectors are summed elementwise rather than multiplied, and the aligned weights become larger as the additive attention achieves higher accuracy than the multiplicative attention. The resulting vector passes through ReLU activation, convolution and GN, and then a sigmoid layer to bring the values into the range of [0,1]. Sequential usage of Softmax gives sparse activation and is preferred. Moreover, the attention gate filters neuron activation in forwards pass as well as in backwards pass, resulting in down weighting of the gradients obtained from the background \cite{oktay2018attention}. After the sigmoid function, the coefficient is up sampled and element wise multiplied with the input feature, which scales the input vector according to the relevance. If we represent the convolution operation with (W$_{\text{g}}$, W$_{\text{f}}$), GN with (G$_{\text{g}}$, G$_{\text{f}}$) and ReLU ($\sigma_{1}(x) = max(0,x))$, then the attention coefficient can be obtained by the following equation, which is presented in Fig. \ref{Figure:4}.

\begin{equation}
F = \sigma_{1}[(W_f^T \times  f + G_{f}) + (W_g^T \times  g + G_{g})] \label{eq} \end{equation}
\begin{equation} \alpha = \sigma_{2}(W_\theta^T \times  F + G_{\theta})  \label{eq} \end{equation}

where W$_{\text{$\theta$}}$ and  G$_{\text{$\theta$}}$  are the output of convolutional and group normalization and $\sigma_{2} = \frac{1}{1 + e^{(-x)}}$.

\subsubsection{Deep supervision}
Deep supervision is a commonly used method to avoid vanishing and exploding gradient problems. This forces the intermediate layers to obtain more discriminative features. Deep supervision was first introduced in \cite{lee2015deeply} and applied in medical image applications \cite{folle2019dilated}. Some applications use deep supervision by downsampling the ground truth and incorporating the loss function by assigning weights to each coefficient. We attempt to aggregate the features of the intermediate layer, which are coarse and have low resolution, in the final layer with a single loss function and then apply Softmax for the output of the segmentation maps, as illustrated in Fig. \ref{Figure:3}. In summary, vanilla UNet has plain skip connections, whereas we proposed the attention gate with group normalization, which can activate certain areas “where” the most semantic information is to search and gives less weight to the background. Moreover, the inclusion of deep supervision enhances the convergence capability of the model and usage of features from the intermediate layer and aggregates the contextual information in the final layer, which can result in better performance.

\subsubsection{Loss function}
The cerebrovascular segmentation task is very difficult because it has quite a large class imbalance. Most of the data contain anatomical structures or background classes, and vessels are very sparse and occupy only 1.5\% of the whole brain volume. Considering this, we cannot use a simple loss function that can give equal weights to both false positives (FPs) and false negatives (FNs). To encounter the severe class imbalance problem, the loss function that we used in our training procedure is Tversky loss \cite{salehi2017tversky}. When P and G are the predicted and ground truth binary labels, the Tversky index is given by the following equation:

\begin{equation}
S(P,G;\alpha,\beta) = \frac {|PG|}{|PG|+ \alpha|P/G| +\beta|P/G|}
\end{equation}
where alpha and beta are the penalties for FPs and FNs, respectively. In this way, FNs weigh more than FPs in highly imbalanced data, such as vessels. The mathematical equation for the loss is given below.

\begin{equation}
T(\alpha, \beta) = \frac{\sum_{i=1}^Np_{0i}g_{0i}} {\sum_{i=1}^Np_{0i}g_{0i} +\alpha\sum_{i=1}^Np_{0i}g_{1i}+\beta\sum_{i=1}^Np_{1i}g_{0i}}
\end{equation}

On the top of the Softmax layer, p$_{\text{0i}}$ is the probability of the $i^{\text{th}}$ voxel of the vessel, and p$_{\text{1i}}$ is the background probability. Similarly, g$_{\text{0i}}$ is 1 for vessels and 0 for the background, and vice versa for g$_{\text{1i}}$.

\section{Experimental Results and Discussion}
\label{results}
\subsection{Dataset}
The dataset used is available publicly and is a dataset of 3D TOF-MRA images and T$_{\text{1}}$-weighted and T$_{\text{2}}$-weighted images collected from 110 healthy patients (\url{https://public.kitware.com/Wiki/TubeTK/Data}). Labels are available for TOF-MRA data in a spatial format for 42 subjects. We converted the spatial objects to NIfTI format by using the open-source toolkit TubeTK \cite{aylward2002initialization}. The dataset was randomly split into 36, 3, and 3 subjects for training, validating, and testing the model, respectively. The dimensions of the 3D TOF-MRA image were $448\times448\times128$ with a voxel size spacing of $0.5134mm\times0.51234mm\times0.8mm$. For the remaining 68 subjects, data labels were generated, and the same preprocessing and patch generation methods were applied and fed to the network. This dataset was randomly split into 56, 6, and 6 subjects for training, validating, and testing the model, respectively.

\begin{table}[]
\centering
\caption{Computational complexity of different models on the TubeTk dataset}
\label{Table:1}
\normalsize
\resizebox{0.47\textwidth}{!}{%
\begin{tabular}{@{}ccc@{}}
\hline\hline
Model            & Parameters (millions) & Inference Time (sec) \\
\hline
3DUnet           & 2.466                 & 5.26                 \\
Vnet             & 45.64                 & 15.22                \\
RE-NET           & 22.57                 & 10.33                \\
Uception         & 9.22                  & 7.31                 \\
CS$^2$-NET          & 5.96                  & 6.17                 \\
CV-AttentionUNet & 24.83                 & 10.94          \\\hline\hline  

\end{tabular} %
}
\end{table}
\subsection{Training procedure}
As described earlier, we used the patch-based technique to train the models. We used the sliding window method to slide the image with an appropriate stride factor. During training, from each MRA image, we obtained 190 patches with dimensions of $64\times64\times64$, and for validation, 50 patches were extracted with a patch size of $96\times96\times96$. Finally, the testing results were obtained using full volume brain images. Similarly, we trained the model on our labeled images. We used the same model and weights and applied transfer learning, and after fine-tuning for 30 epochs, we report the results on our labeled images separately. The detailed training, validation, and testing procedure is depicted below in Fig. \ref{Figure:5}. The initial learning rate (LR) for CV-AttentionNet was 0.0001, and it dropped by 0.1 LR if the model reached a plateau after 10 epochs. We saved the model with the best validated DSC score. The optimizer is Adam, and the minibatch size is 8. We first trained the model on the TTKL dataset for 70 epochs, and after its convergence, we transferred the weights to TTKU\_L data and ran the model for 30 more epochs. It took 12 hours on 4 NVIDIA GeForce TITAN Xp with 12 Gb of memory. The whole model is trained using the PyTorch (\url{https://pytorch.org/}) framework, and the labeled ground truths will soon be available online. The computational complexity of the model with inference time per volume is given in Table \ref{Table:1}.

\begin{figure}[]
  \centering
  \centerline{\includegraphics[width=8.7cm]{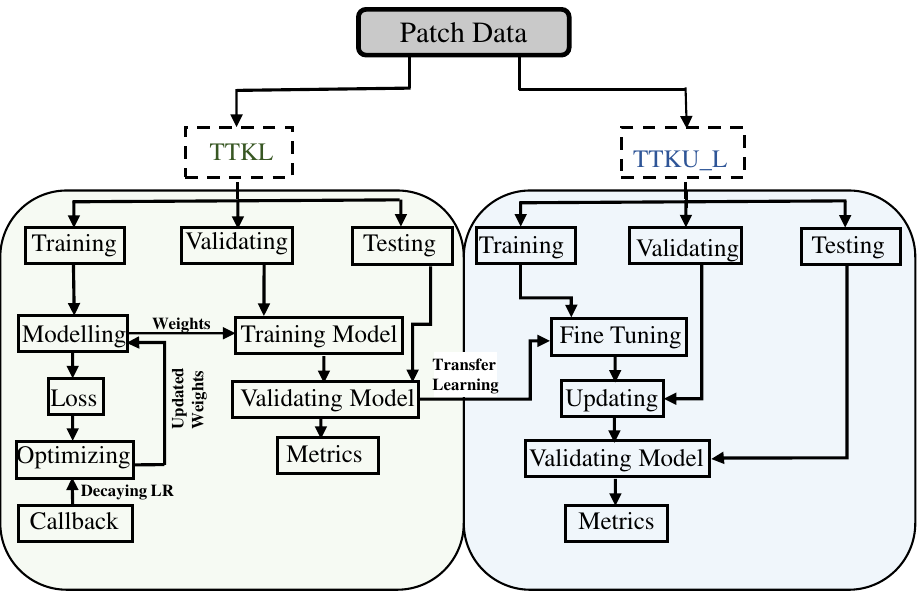}}
  \caption {Detailed training, validation and testing steps.}\label{Figure:5}\medskip
\end{figure}
\subsection{Evaluation}
\subsubsection{Qualitative}
For the quantitative analysis, the authors \cite{moccia2018blood} listed many metrics for different modalities of segmented blood vessels. The DSC score is the measure of overlap between two binary volumes and is defined as follows:

\begin{equation}
          DSC = \frac{2TP}{2TP + FP + FN}
\end{equation}
where TP, FP, and FN are the voxel counts of true positives (correctly predicted), false positives (misclassified as positive), and false negatives (misclassified as negative), respectively. As vessel data are highly imbalanced and most of the volume is the background, we concentrated on the foreground. Therefore, the DSC score is the “correct prediction,” which is defined as the ratio of the foreground of predicted to ground truth ranging from 0 to 1. The positive predictive value is called precision and is defined as follows:

\begin{equation}
          Percision = \frac{TP}{TP + FP}
\end{equation}
It is the ratio of true positives to the entire prediction result. Similarly, sensitivity, which is also called recall, is the ratio of true positives to the whole ground truth given by the following equation.
\begin{equation}
          Sensitivity = \frac{TP}{TP + FN}
\end{equation}
Specificity is a measure of false negatives that are correctly segmented following the equation below.
\begin{equation}
         Specificity = \frac {TN}{TN+FP}
\end{equation}
Last but not least is the distance measure which is Hausdorff Distance (HD) and is described below.
\begin{equation}
          HD = max(sup_{a\epsilon A}inf_{b\epsilon B} d(a,b),sup_{ b\epsilon B}inf_{a\epsilon A} d(a,b)
\end{equation}
where sup is the supremum, inf is the infimum, and d is the absolute distance value.
\begin{figure}[]
  \centering
  \centerline{\includegraphics[width=9cm, height= 3cm]{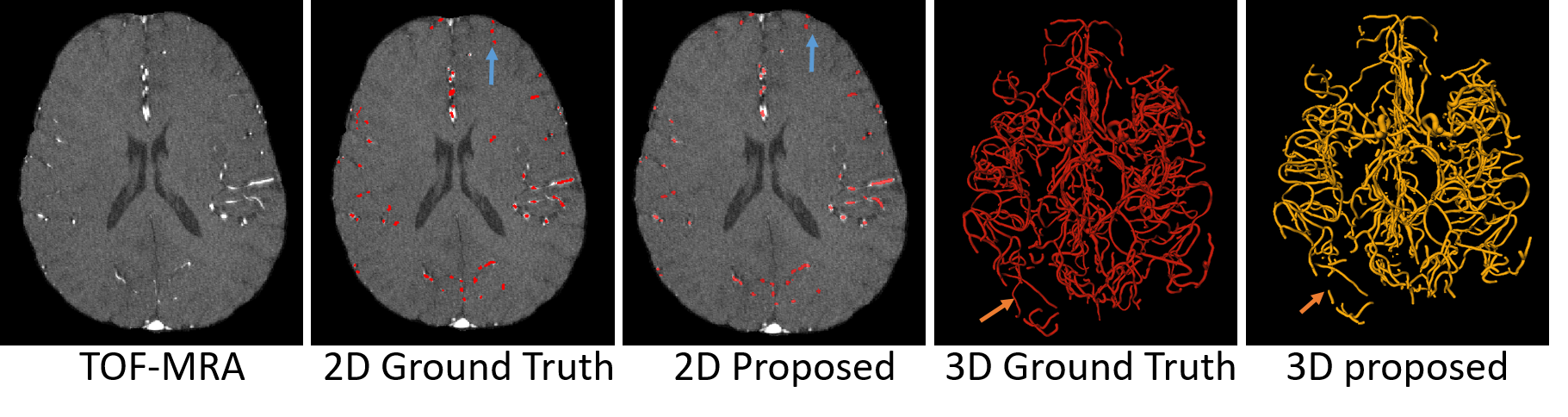}}
  \caption {Qualitative assessment for TubeTK-labeled data: From left to right, the images are the original TOF-MRA image, the image overlaid with the ground truth, the image overlaid with the proposed output, 3D visualization of the ground truth, and 3D visualization of the image using the proposed method. The blue and orange arrows show the missing small vessels in the proposed method.}\label{Figure:6}\medskip
\end{figure}
\begin{figure}[]
  \centering
  \centerline{\includegraphics[width=9cm, height= 3cm]{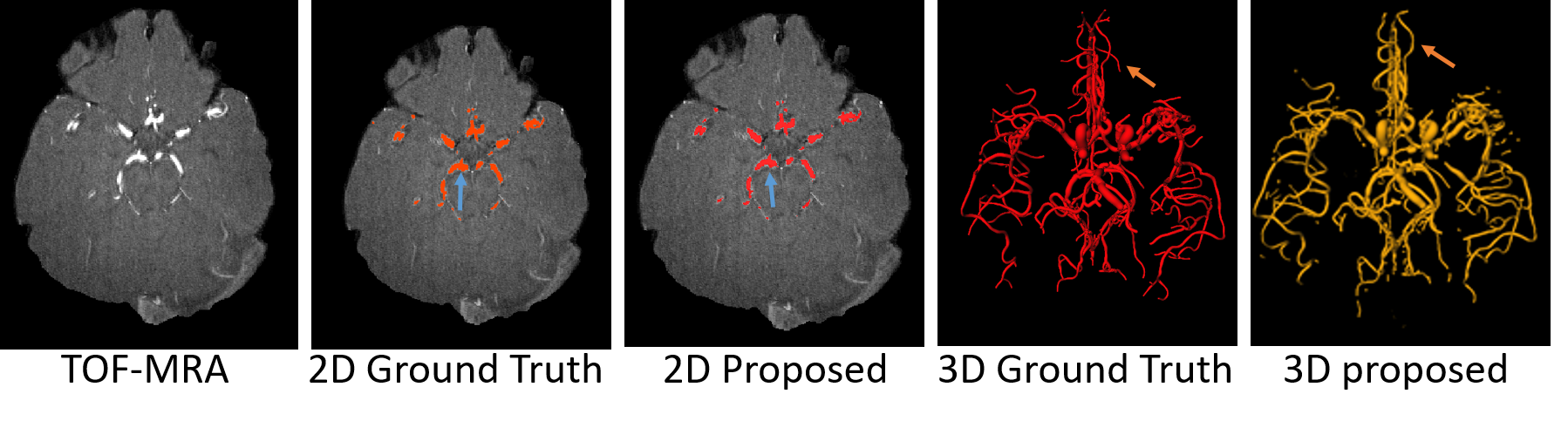}}
  \caption {Qualitative assessment of our labeled data: From left to right, the images are the original TOF-MRA image, the image overlaid with the ground truth, the image overlaid with the output of the proposed method, 3D visualization of the ground truth, and 3D visualization of the image using the proposed method. The blue and orange arrows show the missing small vessels in the proposed method.}\label{Figure:7}\medskip
\end{figure}

\subsubsection{Qualitative}
For the qualitative analysis of prediction masks, they are assessed based on large and small vessels. The large vessels are the internal carotid artery (ICA), anterior cerebral artery (ACA), basilar artery, and M1 segments of the posterior cerebral artery along with P1 segments, as described in \cite{hilbert2020brave}. The rest of the vessels are relatively small vessels. We conducted a range of experiments to evaluate the performance of the proposed method. First, we conducted experiments to evaluate the effectiveness of the enhancement method and proposed an attention module. Second, we compare the results with state-of-the-art methods on the TTKL and TTKU\_L datasets. The details of each are in the following sections.

\subsection{Effectiveness of Enhancement Method}

\begin{table}[]
\centering
\caption{Comparison of the vesselness enhancement method to other enhancement methods}
\label{Table:2}
\normalsize
\resizebox{0.47\textwidth}{!}{%
\begin{tabular}{lcccc}
\hline\hline
\multicolumn{1}{l}{Method}      & \multicolumn{1}{l}{DSC (\%)} & \multicolumn{1}{l}{Precision (\%)} & \multicolumn{1}{l}{Sensitivity (\%)} & \multicolumn{1}{l}{Specificity (\%)} \\ \hline
\rule{0pt}{10pt}
Raw Data                        & 62.75                       & 61.34                             & 63.26                               & 99.85 \\
\rule{0pt}{10pt}
Gamma Correction                & 62.97                       & 61.89                             & 63.75                               & 99.87                               \\
\rule{0pt}{10pt}
Retinex Enhancement             & 64.53                       & 64.05                             & 64.12                               & 99.86                               \\
\rule{0pt}{10pt}
\textbf{Vesselness Enhancement} & \textbf{65.77}              & \textbf{65.37}                    & \textbf{64.85}                      & \textbf{99.87}                      \\ \hline\hline
\end{tabular}%
}

\end{table}
As described earlier, raw MRA images are noisy; therefore, Hessian-based vessel enhancement is used to increase the contrast of small and large vessels while removing the background anatomical structure. To provide a quantitative assessment, we applied other enhancement methods, such as gamma correction and the retinex enhancement method, which have been applied in retinal vessel segmentation. After the enhancements, we trained and tested it using vanilla 3D-UNet, and the results are shown in Table \ref{Table:2}. From Table \ref{Table:2}, it can be observed that there is a significant difference in the DSC score compared to raw, gamma correction and retinex enhancement data. The DSC score is 2.8\% and 1.24\% better in vesselness enhancement compared with the other alternative enhancement methods. Our enhancement method performed better since gamma correction corrects only the luminance value, and the retinex enhancement method removes the luminance effect and reveals the reflectance of the vessel image more objectively, whereas the vesselness enhancement not only enhances vessels of varied sizes but also removes the noisy background anatomical structure. Thus, our enhancement method performs better in terms of all metrics, which leads to significant improvement in the overall segmentation results.

\subsection{Cerebrovascular segmentation using CV-AttentionUNet}
CV-AttentionUNet looks for salient regions, such as vessels, in the image and aggregates the information using deep supervision. As vessels are small and have varied sizes, we have to choose the appropriate patch size. A large patch size in multiple pooling operations may result in making small vessels invisible, and a small patch size may lead to the loss of rich contextual information. Therefore, we chose a $64\times64\times64$ patch volume in accordance with the limitation of computational resources. We randomly picked 190 patches from each volume and obtained 6990 patches for the TTKL dataset. Similarly, for TTKU\_L, we obtained 190 patches of each image, for a total of 11216 patches for training and validation. The quantitative results of our method for TTKL and TTKU\_L data are shown in Table \ref{Table:2}. From the table, it is evident that we achieved significant improvement, with DSC scores of 70.85\% and 91.74\% for both labeled datasets. CV-AttentionUNet achieved remarkable results on cerebrovascular segmentation. The TTKL and TTKU\_L datasets differ in terms of sensitivity, DSC score, and precision. Due to the large number of tiny vessels in the TTKL dataset, almost all of which are the size of a dot, the detection was quite low, as indicated in the sensitivity score in Table \ref{Table:3}. Additionally, they are difficult to segment because of the nature of the data and the errors in conversion from spatial objects to NIfTI format. Similarly, the qualitative results of our method for TTKL and TTKU\_L are shown in Fig. \ref{Figure:6} and Fig.\ref{Figure:7}. It is deduced from both figures that our method can learn more 3D contextual information and can extract the vessel with the help of attention and deep supervision modules. From the qualitative analysis, one can observe that TTKU\_L has similar large vessels, but quite a few small vessels are missing compared to TTKL. In TTKL, some small vessels are missed by the proposed method, as shown by the blue and orange arrows (Fig. \ref{Figure:6}). Similarly, for TTKU\_L, the large and missing vessels are indicated by blue and orange arrows, respectively (Fig. \ref{Figure:7}).


\begin{figure*}[]
  \centering
  \centerline{\includegraphics[width=\textwidth]{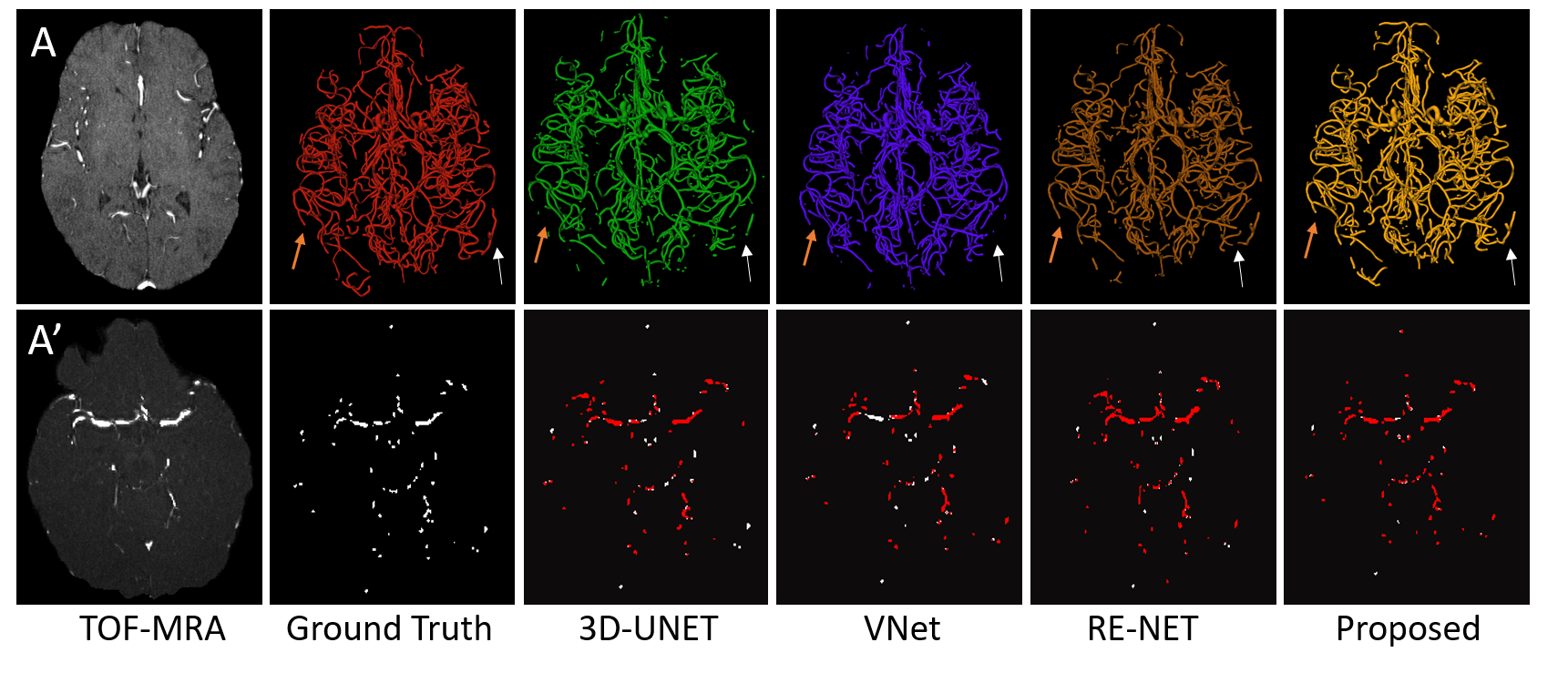}}
  \caption {Qualitative results on the TubeTK-labeled dataset for subject A and its corresponding error map ${\rm{A^\prime}}$. From left to right, the images are the original TOF-MRA, ground truth and output of 3D-UNET, VNet, RE-NET, the proposed method and the corresponding error maps for each model. The comparatively better performance results are highlighted with orange and white arrows. }\label{Figure:8}\medskip
\end{figure*}
\begin{figure*}[]
  \centering
  \centerline{\includegraphics[width=\textwidth]{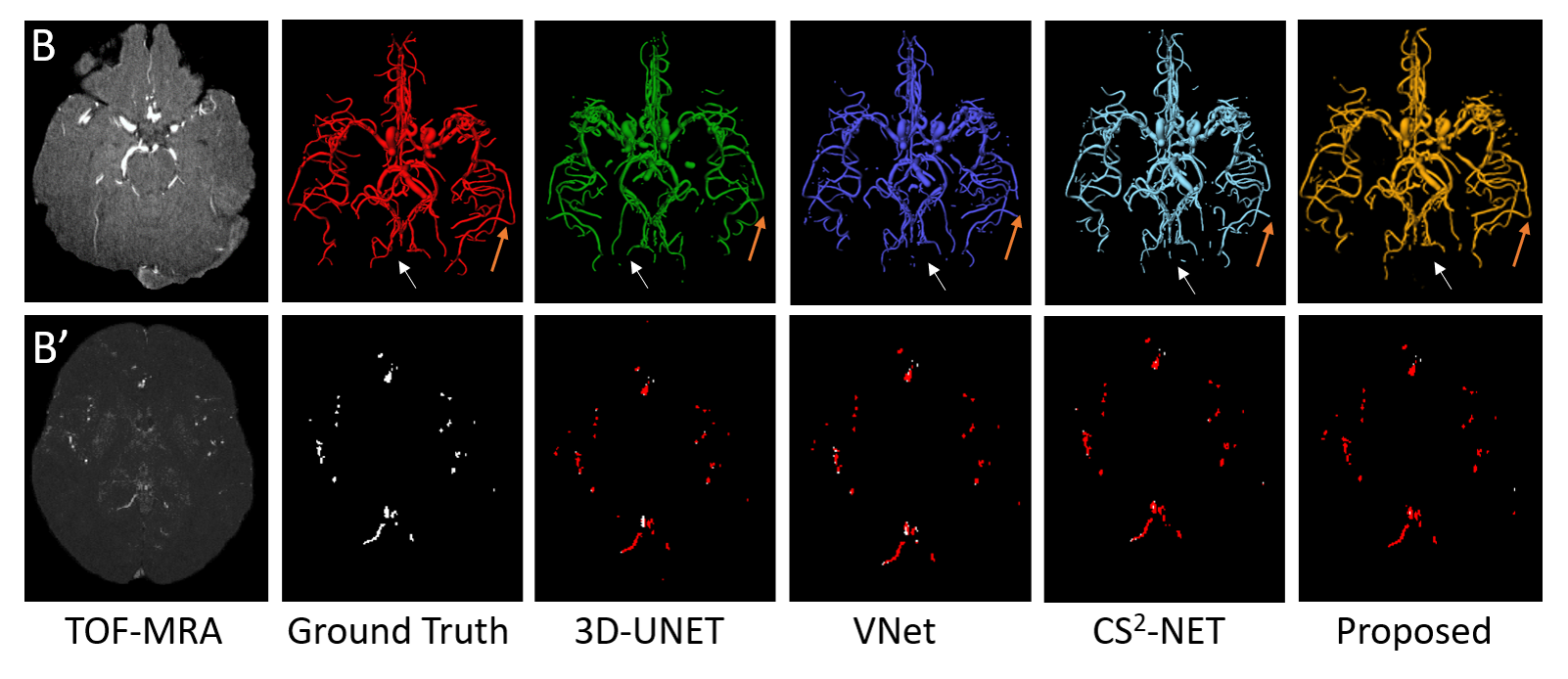}}
  \caption { Qualitative results on our labeled data for subject B and its corresponding error map ${\rm{B^\prime}}$. From left to right, the images are the original TOF-MRA, ground truth and output of 3D-UNET, VNet, CS$^2$-Net, the proposed method and the corresponding error maps for each model. The comparatively better performance results are highlighted with orange and white.}\label{Figure:9}\medskip
\end{figure*}
\begin{table}[]
\centering
\caption{Performance of CV-AttentionUNet on TTKL and TTKU\_L data}
\label{Table:3}
\small
\resizebox{0.47\textwidth}{!}{%
\begin{tabular}{ccc}
\hline\hline
\multicolumn{1}{c}{Metrics} & \multicolumn{1}{l}{TTKL} & \multicolumn{1}{l}{TTKU\_L} \\ \hline
\rule{0pt}{10pt}
DSC (\%)                    & 70.85                    & 91.74                       \\
\rule{0pt}{10pt}
Sensitivity (\%)            & 72.32                    & 93.38                       \\
\rule{0pt}{10pt}
Precision (\%)              & 68.28                    & 90.16                       \\
\rule{0pt}{10pt}
Specificity (\%)            & 99.92                    & 99.97                       \\
\rule{0pt}{10pt}
Hausdroff Distance(mm)      & 1.1152                   & 1.2628                      \\ \hline\hline
\end{tabular}%
}
\end{table}
\begin{table*}[t]
\caption{Your table caption here.}
\label{Table:4}
\resizebox{\textwidth}{!}{%
\begin{tabular}{|l|c|c|c|c|c|c|c|c|c|c|}
\hline\hline
& \multicolumn{5}{c|}{TTKL} & \multicolumn{5}{c|}{TTKU\_L} \\ \cline{2-11} 
& DSC(\%) & Sensitivity(\%) & Precision(\%) & Specificity(\%) & HD(mm) & DSC(\%) & Sensitivity(\%) & Precision(\%) & Specificity(\%) & HD(mm) \\ \hline
3D-UNet \cite{cciccek20163d} & 65.77 & 65.37 & 64.85 & 99.87 & 1.1988 & 88.14 & 86.72 & 89.63 & 99.97 & 1.2748 \\ \hline
V-Net \cite{milletari2016v} & 62.67 & 64.84 & 60.71 & 99.84 & 1.2173 & 90.28 & 91.78 & 88.83 & 99.98 & 1.4512 \\ \hline
Uception \cite{phellan2017vascular} & 67.68 & 65.89 & 66.44 & 99.87 & 1.1501 & - & - & - & - & - \\ \hline
RE-NET \cite{zhang2020cerebrovascular} & 69.90 & 70.70 & 67.87 & 99.88 & \textbf{1.0591} & - & - & - & - & - \\ \hline
CS$^2$-NET \cite{mou2020cs2} & - & - & - & - & - & 90.42 & 87.92 & 91.07 & 99.98 & 1.3175 \\ \hline
\textbf{Proposed} & \textbf{70.85} & \textbf{72.32} & \textbf{68.28} & \textbf{99.92} & 1.1152 & \textbf{91.74} & \textbf{93.38} & \textbf{92.16} & \textbf{99.98} & \textbf{1.2128} \\ \hline\hline
\end{tabular}%
}\end{table*}

\subsection{Comparison with other methods}
The quantitative cerebrovascular segmentation results obtained from several state-of-the-art methods are listed in Table \ref{Table:4} for TTKL and TTKU\_L data. With the TTKL data, CV-AttentionUNet attained the highest performance on all measures, namely, a DSC score of 70.85\%, sensitivity of 72.32\%, specificity of 99.92\%, and precision of 68.28\%, except for HD, and all results are the mean values of three testing subjects. Our method outperforms the other existing methods that have been proposed on the TubeTK dataset except for HD, which is obtained by RE-NET and showed better outcomes. We compared our method with 3D-UNet, VNet  \cite{milletari2016v}, Uception, and RE-NET, and our method showed improvements of 5.08\%, 8.18\%, 3.17\%, and 0.95\%, respectively, in terms of DSC scores. In \cite{zhang2020cerebrovascular}, the DSC score, sensitivity, precision and specificity score were reported to be 69.90\%, 70.70\%, 67.87\% and 99.88\%, respectively, and with our proposed method, we observed improvements of approximately 1\%, 2\%, 1\% and 0.1\%, respectively. Additionally, while comparing our result to \cite{phellan2017vascular}, we witnessed improvements of approximately 3\%, 7\%, 2\%, and 0.1\% from the reported results. Moreover, for the TTKU\_L data, the method achieved a DSC score of 91.74\%, sensitivity of 93.38\%, precision of 92.16\%, specificity of 99.98\%, and HD of 1.2128 mm. The comparative results in terms of the DSC score were 3.6\%, 1.46\%, and 1\% better than those of 3D-UNet, VNet, and CS2-NET, which were averaged over six subjects, respectively.

The qualitative analyses for TTKL and TTKU\_L are shown in Fig. \ref{Figure:8} and Fig. \ref{Figure:9}, indicating UNet and other methods under segments of certain arteries. Large vessels were detected in all approaches, but small vessels were imperceptible. Integration of the enhancement method, particularly the attention module with group normalization, overcomes these limitations and focuses on relevant areas. In our proposed method, false positives are greatly reduced, and small vessels, which are not detected in other methods, are segmented properly in the TTKL data. Unlike other methods, our method was not distracted by the tortuous nature of vessels, and many small vessels were correctly classified.

Particular areas of vessels for emphasis are marked with white, green, and blue arrows in Fig. \ref{Figure:8}, where 3D and 2D cross sections of the same subject are indicated by A and ${\rm{A^\prime}}$, respectively. Similarly, for TTKU\_L, we showed two different cross sections where large vessels were marked with different colors in Fig. \ref{Figure:9} for another subject B. It can be observed that the large vessels are correctly classified with all three comparative methods. However, false positives are present for small vessels, as shown by the orange, white, blue, and green arrows for the two subjects A and B. To draw a comparison between the ground truth and predicted result, we also presented the error maps in Figs. \ref{Figure:8} \& \ref{Figure:9}. For both TTKL and TTKU\_L data, small vessels are detected better when compared with other methods, which shows that our method generalizes data well.

\subsection{Discussion}
By quantitative and qualitative assessment, we can conclude that our proposed method achieved state-of-the-art performance. First, none of the other models, except RE-NET, considered the noncontrasting nature of TOF-MRA images, but their enhancement method did not consider background noise. Second, the vessel dataset is highly imbalanced, and to mitigate this effect, we use Tversky loss, which is attributed to this effect more effectively than the dice loss. Tversky loss penalizes the FPs and FNs, and the inclusion of Tversky loss helps the model generalize well. Third, since the DSC score was insufficient, precision, sensitivity, specificity, and HD were collectively used as quantitative measures. The selection of different metrics provided a sufficient and complete overview of the model performance for vessels. Moreover, qualitative analysis based on brain anatomy, which has not been extensively explored, provides useful insight. Fourth, one of the key developments of our study is the label generation method for cerebrovascular segmentation. We not only defined labels based on the brain anatomy of vessel structures but also compared them with other methods to show the excellent performance of our methodology. Fifth, the superior performance is due to the addition of the attention module with GN. By focusing on salient regions, the attention modules were not confused with redundant data. GN considered the effect of the batch size, which is quite small in our study, while all the other studies did not consider this factor. Finally, the inclusion of deep supervision, which is strong regularization for classification accuracy, learned discriminative features and helped tackle the problematic convergence behavior. It also helped to capture the wider anatomical structures of vessels, particularly small vessels. Furthermore, unmasked brain images can distinguish the brain, skull, and neck areas from other brain areas and vessels.

\section{Conclusion and Future works}
\label{conclusion}
In this work, we proposed an attention-based architecture for cerebrovascular segmentation tasks. We utilized the vessel enhancement filter to balance the contrast and group normalization-based attention module to obtain salient semantic information. The inclusion of deep supervision helped the model converge faster and enabled the utilization of features in intermediate layers. The proposed methodology takes advantage of vessel enhancement, UNet, attention mechanisms, and deep supervision. We compared our approach with several other methods on the TubeTK dataset, corroborating the efficiency of the proposed method. Our approach is an end-to-end approach and provides 3D output, which can help clinicians and surgeons assess the status of cerebrovascular abnormalities before surgeries and treatments. Our study has several limitations. First, we only tested our model on data from healthy patients and not on patients with cerebrovascular disease due to the unavailability of data. Using data on subjects with aneurysms, pretentious fluid, stenosis, and other diseases is highly warranted. Second, in terms of architectural advancement, we only consider spatial attention and neglect channel attention. Finally, to cope with the data deficiency, data augmentation techniques can be included in future studies.

\section*{Acknowledgment}

We sincerely thank the anonymous reviewers for the comprehensive evaluation of our manuscript and Ekta Srivastava (School of Electrical Engineering and Computer Science, Gwangju Insititute of Science and Technology (GIST)) for her support in revising it.


\bibliographystyle{ieeetr}
\bibliography{references}
%

\begin{IEEEbiography}[{\includegraphics[width=1in,height=1.25in,clip,keepaspectratio]{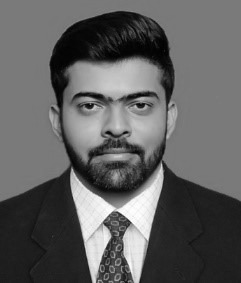}}]{Syed Farhan Abbas} received his B.S. degree in electrical engineering from the University of Gujrat, Gujrat, Punjab, Pakistan in 2016. And received his M.S degree in Biomedical Science and Engineering from Gwangju Institute of Science and Technology, Gwangju, South Korea. Currently, he is working as researcher at Korea University, South Korea. His current research includes medical image processing, deep learning, and its application to medical imaging.
\end{IEEEbiography}

\begin{IEEEbiography}[{\includegraphics[width=1in,height=1.25in,clip,keepaspectratio]{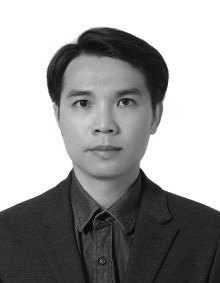}}]{NGUYEN THANH DUC} received a B.S. degree in electrical and electronic engineering from the Vietnam National University- Ho Chi Minh City University of Technology, Ho Chi Minh, Vietnam, in 2010, and his M.Sc. degree from Tampere University of Technology, Tampere, Finland, in 2013. He received his Ph.D. degree from the Department of Biomedical Science and Engineering at Gwangju Institute of Science and Technology, Gwangju, South Korea in 2020. Currently, he is working as AI scientist at Prenuva, Canada. His current research interests include biomedical signal and brain image processing and machine learning applications in neuroimaging.
\end{IEEEbiography}

\begin{IEEEbiography}[{\includegraphics[width=1in,height=1.25in,clip,keepaspectratio]{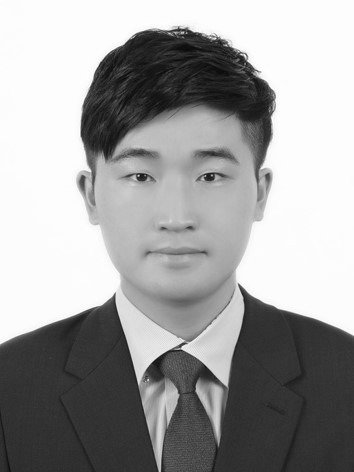}}]{YOONGUU SONG} graduated from Yonsei University, Wonju, South Korea in 2018. Currently, he is a Ph.D. candidate in the Department of Biomedical of Science and Engineering at Gwangju Institute of Science and Technology, Gwangju, South Korea. His research interests include brain image processing such as multimodal registration using deep learning and artificial intelligence applications in neuroimaging.
\end{IEEEbiography}

\begin{IEEEbiography}[{\includegraphics[width=1in,height=1.25in,clip,keepaspectratio]{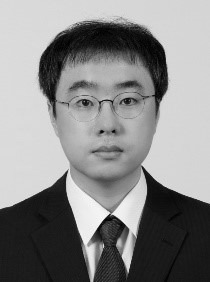}}]{KYUNGWON KIM} graduated from the College of Medicine, Chonnam National University, Gwangju, South Korea and received an M.D. degree in 2012. He received the M.S. degree at the College of Medicine, Pusan National University, in 2016. He completed a residency in psychiatry at Pusan National University Hospital, Busan, South Korea, in 2017. He received his Ph.D. degree in Biomedical of Science and
Engineering at Gwangju Institute of Science and Technology, Gwangju, South Korea. Currently, he is working as clinical assistant professor at Pusan National University, South Korea. His research interests include mental disorders, precision medicine, biomedical signals, and artificial intelligence.
\end{IEEEbiography}

\begin{IEEEbiography}[{\includegraphics[width=1in,height=1.25in,clip,keepaspectratio]{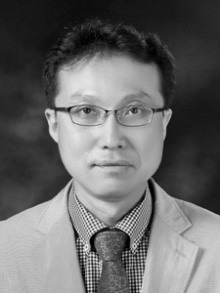}}]{Boreom Lee} graduated from Seoul National University College of Medicine, Seoul, South Korea. He received his M.D. degree in 1998 and his Ph.D. degree in biomedical engineering from Seoul National University in 2007. He joined the faculty of the Gwangju Institute of Science and Technology (GIST) in 2011, where he is currently an associate professor with the Department of Biomedical Science and Engineering (BMSE). His current research interests include brain connectomics, healthcare systems, biomedical signal processing and instrumentation, and machine learning applications in medical research areas.
\end{IEEEbiography}




\end{document}